\documentclass[12pt]{iopart}

\usepackage{iopams}  
\usepackage{latexsym}
\usepackage{graphicx}
\usepackage[utf8]{inputenc} 
\usepackage{graphicx}
\usepackage{dcolumn}
\usepackage{bm}
\begin{document}

\title[Diatomic molecule in a strong infrared laser field]{Diatomic molecule in a strong infrared laser field: level-shifts and bond-length change due to laser-dressed Morse potential}

\author{
S\'{a}ndor Varr\'{o}$^{1,2}$,  
Szabolcs Hack$^{1,3,*}$, 
G\'{a}bor Paragi$^{3,4,5}$,
P\'{e}ter F\"{o}ldi$^{1,3}$,
Imre F. Barna$^{2}$, 
and  Attila Czirj\'{a}k$^{1,3}$ }

\address{
 $^{1}$ELI ALPS, ELI-HU Non-Profit Ltd., Wolfgang Sandner utca 3., Szeged, H-6728, Hungary
 \\
$^{2}$Wigner Research Centre for Physics, Konkoly Thege Mikl\'os \'ut 29 - 33, H-1121 Budapest, Hungary 
\\
 $^{3}$ Department of Theoretical Physics, University of Szeged, Tisza L. krt. 84 - 86, H-6720 Szeged, Hungary 
\\
$^{4}$ Institute of Physics, University of P\'{e}cs, Ifj\'us\'{a}g \'utja
 6, H-7624 P\'{e}cs, Hungary
 \\
 $^{5}$ Department of Medical Chemistry, University of Szeged , D\'om Square 8, H-6720 Szeged, Hungary 
 \\
 $^{*}$ corresponding author: szabolcs.hack@eli-alps.hu
}

\begin{abstract}

We present a general mathematical procedure to handle interactions described by a Morse potential in the presence of a strong harmonic excitation.
We account for permanent and  field-induced terms and their gradients in the dipole moment function, and we derive analytic formulae for the bond-length change and for the shifted energy eigenvalues of the vibrations, by using the Kramers-Henneberger frame. We apply these results to the important cases of $\mathrm{H}_{2}$  and LiH, driven by a near- or mid-infrared laser in the $10^{13}$ $\mathrm{ W/cm^2}$ intensity range. 
\end{abstract}

%
\vspace{2pc}

\section{Introduction}

Diatomic molecules driven by strong laser pulses show  a rich variety of fundamentally important processes,
depending on how the laser pulse parameters are related to the diatomic's properties \cite{FMartin_review_2015,Vibok_LICI_2015,FMartin_review_2017}.
Strong infrared (IR) pulses, typically in the intensity range of $10^{12}-10^{14}$ $\mathrm{ W/cm^2}$, have gained increasing importance in several recent developments with diatomics.  
For certain strong-field scenarios, like e.g. gas HHG, mid-IR laser pulses have considerable advantages over near-IR or visible pulses, see the excellent experimental results in \cite{Carrera_PRA_2006,Goulielmakis_Science_2008,Chini_NatPhoton_2014, Gaumnitz_Opt_Exp_2017}. 
Established pump-and-probe experiments in attosecond and strong-field physics \cite{Krausz_Ivanov_2009_RMP} involve a relatively strong IR pulse, like e.g. diatomic molecules driven by a strong infrared laser field in attosecond streaking experiments \cite{Itatani_PRL_2002,Kitzler_PRL_2002,Eckle_NatPhys_2008, Wei_PRA_2018}.
For diatomic molecules, the change of the internuclear distance caused by a laser pulse has a significant influence on the attosecond streaking spectrum \cite{Wang_PRA_2019} as well as on the formation of vibrational wave packets \cite{Ergler_lochfrass_2006,Fang_lochfrass_2008}.
In general, ionization-delay measurements are valuable tools to investigate the electronic properties of molecules \cite{Biswas_NatPhys_2020}.
An emerging topic in this field of research is  molecules in strong IR fields without ionization \cite{tomatto}, 
having obvious relevance also to important experimental techniques of laser manipulation of molecules, like trapping \cite{fried}, alignment \cite{Mousumi_JCC_2022}, orientation \cite{sidem}.
Ultra-high intensity laser pulses ($10^{18}-10^{20}$ $ \mathrm{ W/cm^2}$) have a pedestal \cite{Kapteyn_OptLett_91}  typically in the intensity range of $10^{10}-10^{14}$ $ \mathrm{ W/cm^2}$ (having also an interesting nontrivial photon statistics \cite{Varro_NPJ_2022}) 
which raises questions and concerns about the true "initial" state of the target molecule when the actual ultra-high intensity pulse reaches it. 

Although the theoretical framework for molecules interacting with laser fields is well established \cite{bandrauk,sindelka}, analytic results are rare and usually approximate in the strong-field domain, while most of the numerical methods \cite{FMartin_review_2015,Majorosi2016,majorosi2018densitymodel,Majorosi_PRA_2020} have heavily increasing cost for IR wavelength. 
Thus, new analytic methods for molecules in strong fields are of great significance.

In this paper, we address the problem of a diatomic molecule driven by a strong IR pulse without ionization. We 
investigate a scenario where electronic transitions are also avoided, thus the long quasi-monochromatic IR laser pulse affects the nuclear vibrations only, neglecting also the rotational degrees of freedom. Vibrations of a diatomic molecule are well described by the Morse potential \cite{morse_1929,terhaar} which has a large literature from textbooks to current research papers, including a number of modifications and improvements \cite{varshni,tietz,hua}. There are also recent treatments based on supersymmetry \cite{Benedict_Morse_99,foldi_epl,foldi_phyreva}. 
Some of the rare systematic studies about the capabilities of the Morse potential and its improved versions  to reproduce experimentally measured properties of diatomics conclude that for certain cases the Morse potential is still the best choice.
E.g., the thermal properties such as heat capacity, enthalpy and entropy of three diatomic molecules, $\mathrm{H}_2$, CO, and $\mathrm{I}_2$ were evaluated with the help of the Morse potential and were compared to experimental data \cite{Khordad_IJMP_2022}. The calculated enthalpy of CO and $\mathrm{I}_2$, based on the Morse potential, are in excellent agreement with experimental data in a wide temperature range.
The accuracy of the Morse, Manning-Rosen, Rosen-Morse and Modified-Rosen-Morse potentials was examined and compared for 34 states of 17 diatomic molecules in Ref. \cite{Pingak_RiC_2021}. For the goodness-of-fit to experimental data, the classical Morse potential is more accurate than the other potentials for certain ground and excited states.
For the ground state of CO and SiS molecules, the Morse potential is the most accurate.

Laser driven nuclear dynamics of a diatomic molecule were investigated earlier by many authors, see e.g. \cite{chu1981,ting_1994,yuan_1998,foldi_epl,Murphy_NJP_2007,Triana_NJP_2022}, however, usually neglecting field-induced terms of the dipole moment function \cite{Gready_1977,Gready_1977_2,Gready_1977_3}. 
Due to the moderately strong-field regime we are interested in, we account for permanent and  field-induced terms and their gradients in the dipole moment function in the present work. We use atomic units, unless otherwise stated.

\section{Theoretical model}

The Hamiltonian of a diatomic molecule in center of mass (COM) coordinate system, in the 
presence of an infrared laser field, treated in dipole approximation and length gauge, is the following:
\begin{equation}
\mathrm{\hat{H}=\hat{H}_{0}+\hat{T}_{n}} - \boldsymbol{\hat{\mu}} \mathbf{F}(t),
\label{molecular_Hamiltonian}
\end{equation}
where $\mathrm{\hat{H}_{0}}$, the so-called electronic Hamiltonian, contains the electronic kinetic
energy operator and the electron-electron, electron-nucleus, nucleus-nucleus potential energy 
operators and $\mathrm{\hat{T}_{n}}$ is the nuclear kinetic energy operator. In the interaction 
term  $ \boldsymbol{\hat{\mu}} $ and $\mathbf{F}(t)$ are the dipole moment operator and the electric field 
strength, respectively.
Assuming laser frequencies that are much smaller than the characteristic frequency of the 
electronic transitions between ground and excited electronic states (e.g. IR laser fields) 
we can simultaneously apply the quasistatic approximation to the electronic dynamics and the 
Born-Oppenheimer (BO) approximation \cite{BornOppenheimerApp}. For any instantaneous value of the electric field we consider 
the solution of the time-independent Schr\"{o}dinger equation (TISE) for the  electronic ground
state $\phi_{0}\left(\mathbf{r};\mathbf{R},t\right)$  to be known:
\begin{equation}
\left[\mathrm{\hat{H}_{0}}-\boldsymbol{\hat{\mu}} \mathbf{F}(t)\right] \phi_{0}\left(\mathbf{r};\mathbf{R},t\right)=
\varepsilon_{0}(\mathbf{R},F(t)) \phi_{0}\left(\mathbf{r};\mathbf{R},t\right),
\end{equation}
where $\mathbf{r}$ and $\mathbf{R}$ stand for all electronic and
 for all nuclear coordinates, respectively. 
 Regarding the dynamics of the electronic system, we consider the time and the nuclear coordinates as parameters.
 
We also assume that $\boldsymbol{\hat{\mu}}$  and $\mathbf{F}(t)$ are parallel. 
Note that for a diatomic gas sample having sufficiently low temperature, we may surely neglect rotations, typically below the 10 K range, which can readily met e.g. in supersonic jet expansion. (The smaller the molecule's moment of inertia is, the higher this temperature is, e.g. for $\mathrm{H}_{2}$, below ca. 100 K rotational degrees of freedom are already frozen in.) 
Then the laser field itself can align the molecules with its polarization directions via the induced molecular dipole moment, since requirements for the adiabatic nonresonant molecular alignment \cite{Larsen_Stapelfeldt_JChemPhys_1999} can be fulfilled.
 
Thus, instead of $\mathbf{R}$, we use the internuclear distance $R$ for the diatomic molecule. The field-dependent electronic ground state energy, which plays the role of a potential surface for the nuclear motion, is the sum of two terms:
\begin{equation}
\varepsilon_{0}(R,F(t))=\varepsilon_{0}(R,0)-\intop_{0}^{F\left(t\right)}\mu\left(R,F'\right)\mathrm{d}F',
\end{equation}
where $ \varepsilon_{0}(R,0) $ can be approximated by the Morse potential and $ \mu(R,F) $ is the dipole moment function which includes the field-free and induced terms.

Applying the BO approximation and projecting the molecular Schr\"{o}dinger equation onto $\phi_{0}\left(\mathbf{r};R,t\right)$, the time-dependent Schr\"{o}dinger equation (TDSE) for the nuclear motion becomes:
\begin{equation}
i\hbar\frac{\partial}{\partial t}\chi(u,t)=\left[-\frac{\hbar^{2}}{2 M_{n}}\frac{\partial^{2}}{\partial u^{2}}+V_{M}(u)-\intop_{0}^{F\left(t\right)}\mu\left(u,F'\right)\mathrm{d}F'\right]\chi(u,t),\label{mol_Schrodinger}
\end{equation}
where $\chi\left(u,t\right)$ represent the nuclear wave function, $ M_{n} $ is the reduced nuclear mass,  $u=R-R_{0}$  measures the distance of the two nuclei from its equilibrium value $ R_{0} $. The $ V_{M}(u) $ is the Morse potential with the form of
\begin{equation}
V_{M}(u)=D\left(e^{-2au}-2e^{-au}\right),
\label{morse}
\end{equation}
where  $D>0$ is the potential strength or dissociation energy and $a$ determines the width of the potential well. We take into account the dipole moment function in the following form:
\begin{equation}
\mu(u,F(t))=\mu_{0}+\mu_{1}u+\alpha_{0}F(t)+\alpha_{1} u F(t),
\label{dipole_moment}
\end{equation}
where $ \mu_{0} $ and $ \mu_{1} $ are the permanent dipole moment and dipole gradient,  $ \alpha_{0} $ and $ \alpha_{1} $ are the polarizability and polarizability gradient of the molecule, respectively.

\section{Analytic results for level shifts and bond-length change in the case of  a monochromatic laser excitation }

In the following we assume a long monochromatic laser pulse, i.e. a harmonic excitation $ F(t)=F_{0}\sin(\omega t) $ with peak electric field strength  $ F_{0} $ and angular frequency $ \omega $.

We can easily eliminate the terms containing $ \mu_{0} $ and $ \alpha_{0} $  in  \eref{mol_Schrodinger} with the following unitary transformation:
\begin{equation}
\chi(u,t)=\exp\left\{\frac{i}{\hbar} \intop_{-\infty}^{t} \mu_{0} F(t')+\alpha_{0} F(t')^{2} \mathrm{d}t' \right\}\varphi_{0}(u,t),
\label{unitary_1}
\end{equation}
and we get
\begin{equation}
i\hbar\frac{\partial}{\partial t}\varphi_{0}(u,t)=\left[-\frac{\hbar^{2}}{2 M_{n}}\frac{\partial^{2}}{\partial u^{2}}+V_{M}(u)+ug(t)-\frac{\alpha_{1}F_{0}^{2}}{4}u\right]\varphi_{0}(u,t),
\label{schrodinger_0}
\end{equation}
where $ g(t)= F_{0}[-\mu_{1} \sin(\omega t)+\frac{\alpha_{1}}{4}F_{0}\cos(2\omega t)]$. We apply the next unitary transformation, which is similar to the ${\bf p}\cdot{\bf A}$ gauge transformation:
\begin{equation}
\varphi_{0}(u,t)=\exp\left\{\frac{i}{\hbar} u G(t) \right\}\varphi_{1}(u,t),\;\mathrm{where}
\;G\left(t\right)=\intop_{-\infty}^{t}g\left(t'\right)\mathrm{d}t'.
\end{equation}
 Then a  term with $G(t)^{2}$ appears in the TDSE, which we  eliminate by the unitary  transformation 
  $\varphi_{1}\left(u,t\right)=\exp\left\{ \frac{i}{\hbar}\intop_{-\infty}^{t}G^{2}\left(t'\right)\mathrm{d}t'\right\} \varphi_{2}\left(u,t\right)$. 
 Now the TDSE takes the following form:
\begin{equation}
i\hbar\frac{\partial}{\partial t}\varphi_{2}(u,t)=\left[\frac{\hat{p}^{2}}{2 M_{n}}-\frac{\hat{p}}{M_{n}}G(t) +V_{M}(u)-\frac{\alpha_{1}F_{0}^{2}}{4}u \right]\varphi_{2}(u,t),
\label{schrodinger_2}
\end{equation}
Within the coordinate range of the bound states, the term $V_{M}(u)-\frac{\alpha_{1}F_{0}^{2}}{4}u $  can be very well approximated again by an other Morse potential $ \mathcal{V}_{M}(u) $ having the parameters $\tilde{a}$ and $\tilde{D}$:   
by comparing the Taylor series of $ V_{M}(u)-\frac{\alpha_{1}F_{0}^{2}}{4}u $ to that of $ \mathcal{V}_{M}(u) $  up to the cubic term, 
the $\tilde{a}$ and $\tilde{D}$  are defined by the following equations:
\begin{eqnarray}
2\tilde{a}^{2}\tilde{D}\Delta R_{0}=\frac{\alpha_{1}F_{0}^{2}}{4},\nonumber 
\\
\tilde{a}^{2}\tilde{D}\left( 1+3\tilde{a}\Delta R_{0} \right)=a^2 D,\nonumber 
\\
\frac{1}{3}\tilde{a}^{3}\tilde{D}\left( 3+7\tilde{a}\Delta R_{0} \right)=a^3 D,
\label{analytic_fitting}
\end{eqnarray}
where $\Delta R_{0}$ is the shift of the equilibrium position, i.e. a bond-length change caused by the laser field.

At this point we go over to the so-called Kramers-Henneberger frame \cite{kramers,henneberger,varro_integral_eq,Varro_2019} with
\begin{equation}
\varphi_{2}(u,t)=\exp\left\{\frac{i}{\hbar}\hat{p} \Lambda(t) \right\}\varphi_{3}(u,t),
\;\mathrm{where}
\;\Lambda\left(t\right)=\frac{1}{M_{n}}\intop_{-\infty}^{t}G\left(t'\right)\mathrm{d}t'.
\end{equation}
In this frame the potential energy function is shifted by the following oscillating term:
\begin{equation}
\Lambda(t)=c_{1}\sin(\omega t)+ c_{2}\cos(2\omega t),
\end{equation}
where $c_{1}=\mu_{1}F_{0}/M_{n} \omega^2$ and  $ c_{2}=\alpha_{1}F_{0}^2/ 16 M_{n} \omega^2$.
The explicit form of the shifted potential with the modified $\tilde{D}$ and $\tilde{a}$ parameters is 
\begin{equation}
\mathcal{V}_{M}(u+\Lambda(t))=\tilde{D}\left(e^{-2\tilde{a}(u+\Lambda(t))}-e^{-\tilde{a}(u+\Lambda(t))}\right),
\end{equation}
which contains all of the high-harmonics of the laser field according to series expansions:
\begin{eqnarray}
\exp\left\{-\tilde{a}c_{1}\sin(\omega t)\right\}&=I_{0}(\tilde{a}c_{1})+2\sum_{k=1}^{\infty}I_{k}(\tilde{a}c_{1})\cos[k(\omega t+\pi/2)],\nonumber \\
\exp\left\{-\tilde{a}c_{2}\cos(2\omega t)\right\}&=I_{0}(\tilde{a}c_{2})+2\sum_{k=1}^{\infty}I_{k}(\tilde{a}c_{2})\cos[2k\omega t],
\end{eqnarray}
where $I_{k}(x)$ is the modified Bessel function of the first kind. 
If the laser frequency sufficiently higher than the vibrational and transition frequencies,
then  the nuclei "feel" only the time average of the shifted potential over one optical cycle:
\begin{eqnarray}
\fl \langle \mathcal{V}_{M}(u+\alpha(t))\rangle =\tilde{D}  &\left[
 e^{-2\tilde{a}u} \left(I_{0}(2\tilde{a}c_{1})I_{0}(2\tilde{a}c_{2})+2\sum_{k=1}^{\infty}(-1)^k I_{2k}(2\tilde{a}c_{1})I_{k}(2\tilde{a}c_{2})\right) \nonumber \right. \\
&\left. -2e^{-\tilde{a}u} \left(I_{0}(\tilde{a}c_{1})I_{0}(\tilde{a}c_{2})+2\sum_{k=1}^{\infty}(-1)^k I_{2k}(\tilde{a}c_{1})I_{k}(\tilde{a}c_{2})\right) \right],
\end{eqnarray}
in which the infinite sums are negligible because the arguments of the modified Bessel functions are very small in the reasonable parameter range. Thus, we get a Morse potential with field modified parameters in Kramers-Henneberger frame \cite{kramers,henneberger}:
\begin{equation}
\overline{\mathcal{V}}_{M}(u)=\left\langle \mathcal{V}_{M}(u+\alpha(t))\right\rangle=\overline{D}\left(e^{-2\tilde{a}(u-\overline{u}_{0})}-2e^{-\tilde{a}(u-\overline{u}_{0})}\right),\label{morse_KH}
\end{equation}
where 
\begin{equation}
\overline{D}\equiv \tilde{D}\frac{I_{0}^{2}(\tilde{a}c_{1})I_{0}^{2}(\tilde{a}c_{2})}{I_{0}(2\tilde{a}c_{1})I_{0}(2\tilde{a}c_{2})},
\;\mathrm{and}
\;\overline{u}_{0}\equiv\frac{1}{\tilde{a}}\log\left[\frac{I_{0}(2\tilde{a}c_{1})I_{0}(2\tilde{a}c_{2})}{I_{0}(\tilde{a}c_{1})I_{0}(\tilde{a}c_{2})}\right],
\label{shifted_parameters}
\end{equation}
 are the dissociation energy and the equilibrium position of the Morse potential in the laser field, respectively.
Finally, we obtain the energy eigenvalues of \eref{mol_Schrodinger} as
\begin{equation}
\mathcal{E}_{n}=-\overline{D}+\hbar\overline{\omega}\left(n+\frac{1}{2}\right)-\frac{\hbar^{2}\overline{\omega}^{2}}{4\overline{D}}\left(n+\frac{1}{2}\right)^{2},
\label{shifted_energy}
\end{equation}
where $\overline{\omega}=\tilde{a}\left(\frac{2\overline{D}}{M_{n}}\right)^{1/2}$.

\section{Vibration level shifts of LiH and $\mathrm{H}_{2}$}

We apply our model to LiH and to $\mathrm{H}_{2}$ as examples of heteronuclear and homonuclear molecules, respectively.
 These belong to the most studied diatomic molecules, since both of them have fundamental importance from basic molecular physics to material science
and to astrophysics and cosmology 
\cite{shi,kang,bou,bov2012,wang_2009,Ibrahim_2018}. 

\begin{figure}[p]
\includegraphics[width=0.8\textwidth]{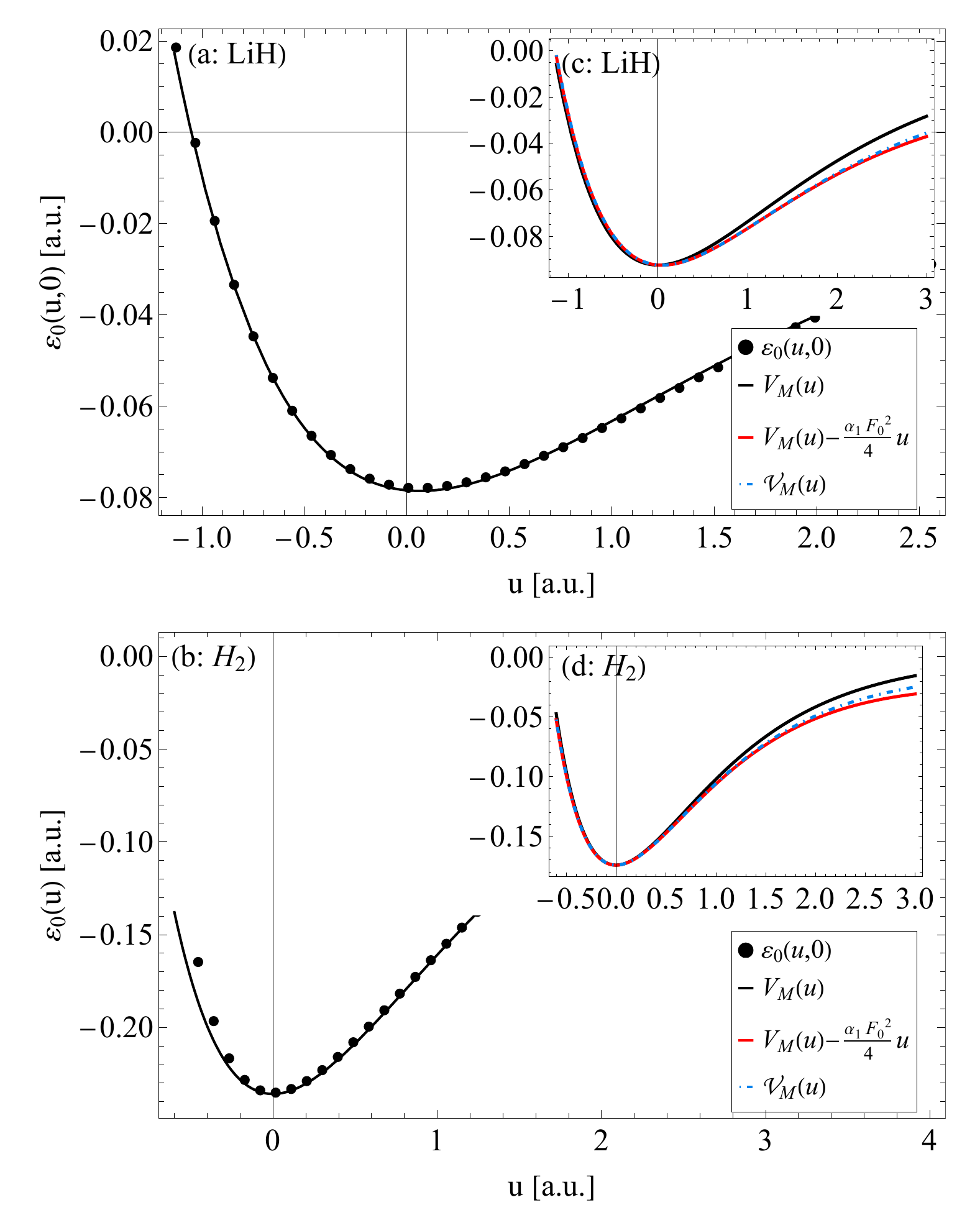}
\protect\caption{Potential energy curves of LiH (upper panel) and $\mathrm{H}_{2}$ (lower panel) according to different models. On (a) and (b) the dots 
give the values of $\varepsilon_{0}(u,0)$ computed by DFT in the indicated range of internuclear separation ($u$), and the solid line plots the Morse potential  $ V_{M}(u)$ fitted to the computed values. The insets (c) and (d) show $ V_{M}(u)$ (black solid line), the term $ V_{M}(u)-\frac{\alpha_{1}F_{0}^{2}}{4}u $ from \eref{schrodinger_2}  (red solid line) and the analytically fitted Morse potential $\mathcal{V}_{M}(u)$ (blue dash-dotted line) according to \eref{analytic_fitting} for $F_{0}=0.02$ a.u. for LiH and $F_{0}=0.04$ a.u. for $\mathrm{H}_{2}$.}
\label{fig:morse_curves}
\end{figure}

\begin{figure}[p]
\includegraphics[width=0.8\textwidth]{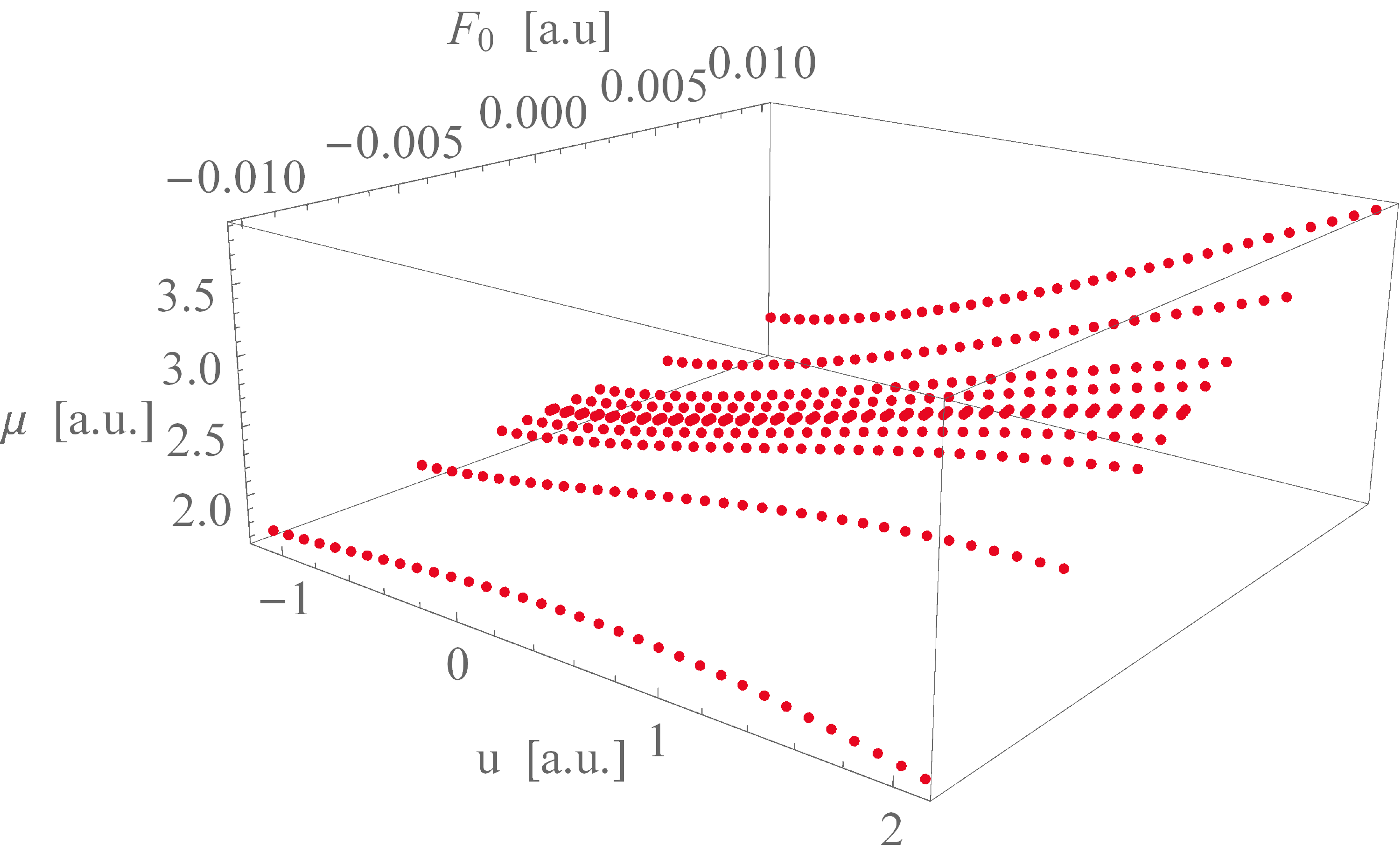}
\protect\caption{Field-dependent dipole moment values $(\mu(u,F(t)))$ of LiH computed by DFT in the indicated range of internuclear separation ($u$) and of static electric field strength ($F_{0}$).}
\label{fig:dipolemoment_function}
\end{figure}

We computed the field-dependent electronic ground state energy $ \varepsilon_{0}(u,F) $ and the molecular dipole moment function $\mu(u,F(t))$ based on density-functional theory (DFT), using 
the Gaussian 16 program package \cite{g16}.
The calculation method was selected according to the suggestion of recent benchmark studies \cite{Zapata_JPC_2020,Hait_JCTC_2018}, where dipole moments for a large test set of molecules, including diatomic molecules were calculated and compared to experimental data. According to the conclusion of Zapata and McKemmish \cite{Zapata_JPC_2020}, double hybrid functionals (e.g. B3LYP \cite{Becke_JCP_1993} ) with quadrupole-$\zeta$-quality basis set \cite{Weigend_PCCP_2005,Weigend_PCCP_2006} is a proper choice for the calculation of small molecules with high accuracy. 
\Fref{fig:morse_curves} (a) and (b) present the computed values of $ \varepsilon_{0}(u,0) $ and it clearly shows that $ \varepsilon_{0}(u,0) $ can be properly approximated for both molecules in the desired coordinate range by a Morse potential.
We present the computed dipole moment function of LiH in  \fref{fig:dipolemoment_function}, which is well fitted by \eref{dipole_moment} in the field strength range from -0.02 to 0.02 a.u. \Tref{table_1} contains the numerical values of the relevant parameters for both LiH and $\mathrm{H}_{2}$.
\begin{center}
\begin{table}[p]
\caption{\label{table_1}The numerical values in a.u. of the parameters of the dipole moment function, \eref{dipole_moment} and of the field-free and the shifted Morse potential for both LiH and $\mathrm{H}_{2}$. The parameters of the laser field are $F_{0}=0.02$ a.u. and $\lambda=1500$ nm.
Numbers in parenthesis are measured values from Ref. \cite{CRC}, numbers in square brackets are computed values from Ref. \cite{Gready_1977_3}.}

\begin{indented}
\lineup
\item[]\begin{tabular}{lllllllllll}
\br
&  $\mu_{0}$ &  $\mu_{1}$ & $\alpha_{0}$ & $\alpha_{1}$ & $D$ & $a$ &  $R_{0}$ & $\overline{D}$ & $\tilde{a}$ & $\overline{R}_{0}$ \\
\mr
 LiH &   2.261  & \00.3673        &  35.72    & \029.16 &  0.09242   &  0.5969   &  3.016  & 0.08194 & 0.6078 & 3.064 \\
        & (2.315) &  [0.373]  & (25.91) &   [22.0]    & (0.08925) & (0.6072) & (3.015) &                &               &             \\ 
\mr
 $\mathrm{H}_{2}$ & 0   & 0 &  5.98         & 12.89 &  0.1744  &  1.028   &   1.401 & 0.2019 & 1.005 & 1.368  \\ 
                                 & (0) &               & (5.41392) &               & (0.1646) & (1.058) & (1.402) &             &             &              \\  
%
%
 \br
\end{tabular}
\end{indented}
\end{table}
\end{center}

 In the case of  $\mathrm{H}_{2}$, since it is a homonuclear molecule, $\mu_{0}$ and $\mu_{1}$ disappear in \eref{dipole_moment}, and $\alpha_{0}$ and $\alpha_{1}$ are the leading and thus the significant terms. Furthermore, \fref{fig:morse_curves} (c) and (d) clearly show that the $ V_{M}(u)-\frac{\alpha_{1}F_{0}^{2}}{4}u $ term in  \eref{schrodinger_0} is really well approachable by Morse potential having different parameters than the original one of the given molecule.
Furthermore, the polarizability gradient more significant for both hetero- and homonuclear molecules, regarding the vibrational level shift. In particular, both for Lih and for $\mathrm{H}_{2}$, the polarizability gradient has a positive sign, thus $ V_{M}(u)-\frac{\alpha_{1}F_{0}^{2}}{4}u $ appears as a less bonding potential, i.e. $\alpha_{1}>0$ causes bond-softening  \cite{Saenz_bondsoftening_PRL_2000}. The above statements are also
      supported by the difference of the number of bound states 
between the original Morse potential, $V_{M}(u)$, and the approximated 
potential, $\overline{\mathcal{V}}_{M}(u)$. For a Morse potential with given 
parameters the number of bonding states is the integer part of the following:
\begin{equation}
N(D,a,M_{n})=\frac{2D-\hbar \omega_{0}}{\hbar \omega_{0}},
\label{number_of_boundstates}
\end{equation}
where $ \omega_{0}$ depends on the parameters of the Morse potential as stated above. In the absence and presence of a laser field having 1500 nm wavelength and  field strength of $0.02$ a.u. for LiH and $0.04$ a.u. for $\mathrm{H}_{2}$, LiH has 27 and 25 bound states, while for $\mathrm{H}_{2}$ these numbers are 16 and 15, respectively.

\begin{figure}[p]
\includegraphics[width=0.8\textwidth]{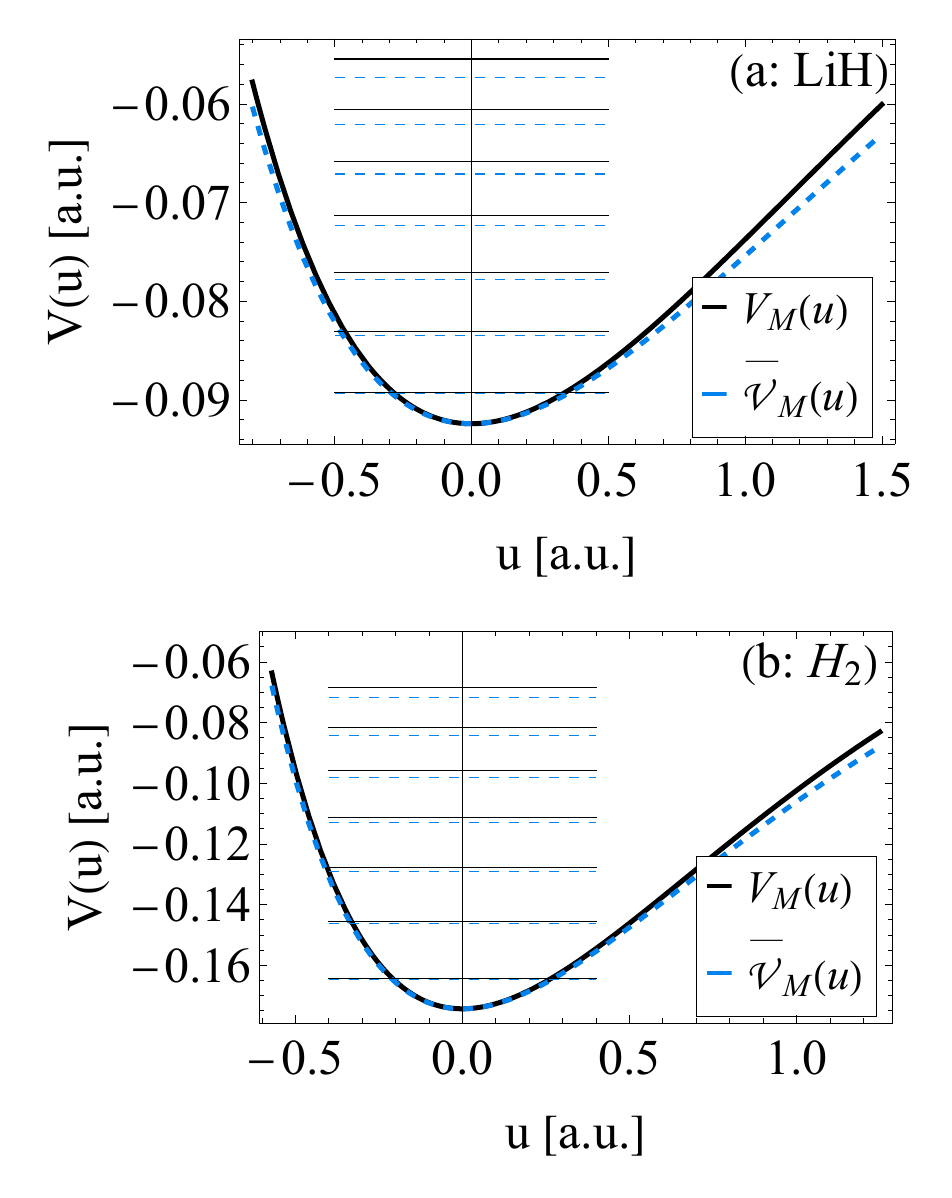}
\protect\caption{The field-free $(V_{M}(u))$ and the shifted Morse potential $(\overline{\mathcal{V}}_{M}(u))$ of LiH (a) and $\mathrm{H}_{2}$ (b) plotted by black solid curve and blue dash-dotted curve, respectively. The horizontal lines with corresponding styles mark the first seven energy levels of the corresponding Morse potential. The parameters of the laser field are $\lambda=1500$ nm and $F_{0}=0.02$ a.u. for LiH, $F_{0}=0.04$ a.u. for $\mathrm{H}_{2}$.}
\label{fig:morse_energy_levels}
\end{figure}
\Fref{fig:morse_energy_levels}  shows how a laser field having the indicated parameters distorts the potential energy curves of LiH (a) and of $\mathrm{H}_{2}$ (b). The fist seven energy levels belonging to  $V_{M}(u)$ and $\overline{\mathcal{V}}_{M}(u)$ are also indicated on \fref{fig:morse_energy_levels} and they clearly show how the laser field shifts the vibrational levels of the molecules. For both of these molecules, as  $\overline{\mathcal{V}}_{M}(u)$ is less bonding potential than $V_{M}(u)$, the spacing between the energy levels are smaller in the presence of the laser field.
\begin{figure}[p]
\includegraphics[width=0.8\textwidth]{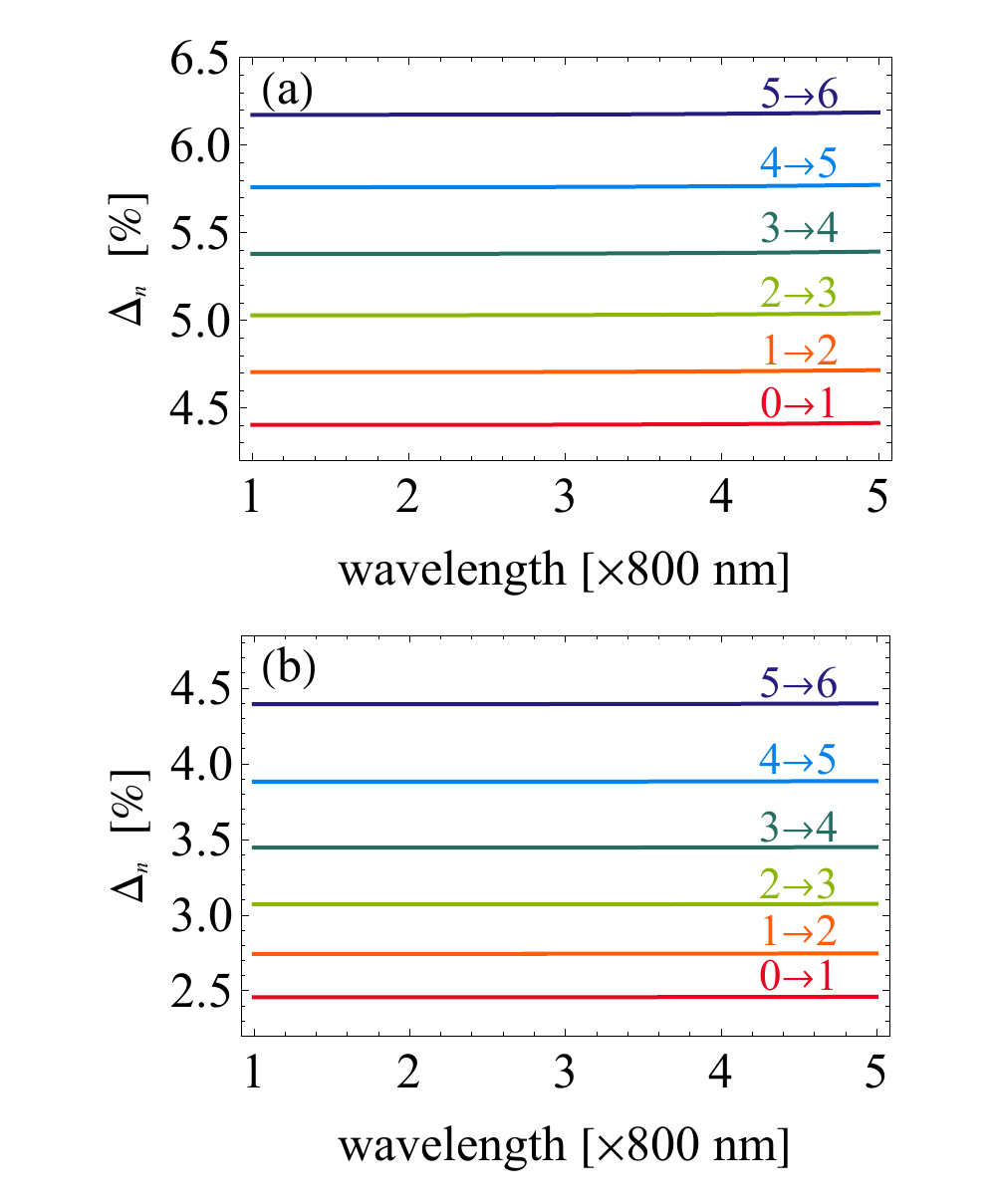}
\protect\caption{Relative shifts of vibrational transition frequency caused by the laser field for the first six energy levels for LiH (a) and for $\mathrm{H}_{2}$ (b) as a function of laser wavelength in case of $F_{0}=0.02$ a.u. for LiH and $F_{0}=0.04$ a.u. for $\mathrm{H}_{2}$.}
\label{fig:morse_rel_shift}
\end{figure}

In order to further investigate the applicability of our method we examine how the spacing of the shifted energy levels varies with the laser wavelength. The relative shift of the transition energy levels of $\overline{\mathcal{V}}_{M}(u)$ compared to the energy level spacing of $V_{M}(u)$ is the following:
\begin{equation}
\Delta_{n}=\frac{|\Delta E_{n}-\Delta \mathcal{E}_{n}|}{\Delta E_{n}},
\label{rel_shift}
\end{equation}
where $\Delta E_{n}=E_{n}-E_{n-1}$ and $\mathcal{E}_{n}=\mathcal{E}_{n}-\mathcal{E}_{n-1}$  are the differences of the energy eigenvalues of $V_{M}(u)$ and of $\overline{\mathcal{V}}_{M}(u)$, respectively. The \eref{rel_shift} quantity in atomic units is actually identical to the relative shifts of vibrational transition frequency. We present the wavelength dependence of $\Delta_{n}$ for the first six vibration transition for both molecules on  \fref{fig:morse_rel_shift}. The shift of the transition energy values caused by the laser field is approximately independent of laser wavelength. This suggests that our model can also be applied to laser pulses in the IR spectral range.

\section{Summary}

We derived analytic expressions, \eref{shifted_parameters} 
and \eref{shifted_energy}, which are valid in a wide wavelength range to account for the effect of a strong long laser pulse on the vibrational levels and the bond-length change of a diatomic molecule with high accuracy.   
These results can be readily applied both to heteronuclear and to homonuclear diatomic molecules and to certain (e.g. alkali metal) atomic dimers. Our model can be relevant also in cases where the Morse potential can be used to describe certain interactions, like weakly bound hydrogen in a larger molecule \cite{Blaise_2000,Zdravkovic_2006}, polyatomic molecule with a single weak internal bond, and the interaction of an atom or a diatomic molecule with a solid surface \cite{Galashev_2019,Safina_2019}.
The resulting level shifts and the corresponding vibrational transition frequency shifts can be readily measured with state of the art spectroscopic methods, both in laboratory and in astrophysical observations. Our results may enable to measure the bond softening and bond length change based on the predicted level shifts in the presence of the laser field.
 

\section{Acknowledgment}
We thank F. Bog\'ar and \'A. Vib\'ok for stimulating discussions. 
The ELI ALPS project (GINOP-2.3.6-15-2015-00001) is supported by the European Union and co-financed by the European Regional Development Fund. One of the authors (G.P.) acknowledges the financial support from the Hungarian Government under project number 2018-1.2.1-NKP-2018-00010.

\section*{References}
\bibliographystyle{iopart-num}
\bibliography{forras_morse}

\end{document}